\documentclass[aps,prl,twocolumn,floats,floatfix]{revtex4}

\usepackage{amsmath}
\usepackage{graphicx}
\usepackage{natbib}
\usepackage{latexsym}

\usepackage{ifthen}
\usepackage{ifpdf}
\ifpdf
\usepackage{graphicx}
\usepackage{epstopdf}
\else
\usepackage{graphicx}
\usepackage{epsfig}
\fi

\newcommand{\ei}{\hat{a}}
\newcommand{\eidag}{\hat{a}^{\dag}}
\newcommand{\hn}{\hat{n}}

\newcommand{\Lx}{\hat{L}_x}
\newcommand{\Ly}{\hat{L}_y}
\newcommand{\Lz}{\hat{L}_z}

\newcommand{\hide}[1]{}

\begin{document}

\title{Interaction-induced dynamical phase locking of Bose-Einstein condensates}

\author{Erez Boukobza$^1$,  Doron Cohen$^2$, and Amichay Vardi$^{1,3}$}
\affiliation{$^1$Department of Chemistry, Ben-Gurion University of the Negev, P.O.B. 653, Beer-Sheva 84105, Israel\\
$^2$Department of Physics, Ben-Gurion University of the Negev, P.O.B. 653, Beer-Sheva 84105, Israel\\
$^3$ITAMP, Harvard-Smithsonian Center for Astrophysics, 60 Garden St., Cambridge Massachusetts 02138}

\begin{abstract}
We show that interactions result in the emergence of a {\it definite} relative-phase between two initially incoherent Bose-Einstein condensates. The many-realization interference fringe visibility is universal at $g_{12}^{(1)}\sim1/3$  throughout the Josephson interaction regime, as evident from a semiclassical picture. Other types of incoherent preparation yield qualitatively different coherence dynamics. 
\end{abstract}


\maketitle

%%%%%%%%%%%%%%%%%%%%%%%%%%%%%%%%%%%%%%%%%%%%%%%%%%%%%%%%%%%%%%%%%%%%%%%%%%%%%%%%%%%%%%%%%%%%%%
%%%%%%%%%%%%%%%%%%%%%%%%%%%%%%%%%%%%%%%%%%%%%%%%%%%%%%%%%%%%%%%%%%%%%%%%%%%%%%%%%%%%%%%%%%%%%%
%%%%%%%%%%%%%%%% INTRODUCTION %%%%%%%%%%%%%%%%%%%%%%%%%%%%%%%%%%%%%%%%%%%%%%%%%%%%%%%%%%%%%%%%
Ever since their first realization \cite{ Andrews97}, interference experiments between two Bose-Einstein condensates (BECs) released from a double-well trap, have  raised fundamental questions concerning gauge symmetry  breaking and the appearance of macroscopic coherence in the bose quantum gas. Two extreme cases are usually contrasted: When the condensates are separated, the state of the system corresponds to a relative-number squeezed state, approaching a {\it Twin-Fock} preparation for fully-independent and equally populated BECs. The corresponding interference pattern was predicted to have an arbitrary relative phase in each experimental run,  varying randomly from one realization to another \cite{Javanainen96,Naraschewski96,Castin97,Rohrl97}. Thus, the many-realization average fringe visibility vanishes for such preparation. By contrast, when the two condensates are  coupled, the initial state corresponds to a {\it coherent} preparation with a {\it definite} relative phase between the constituent BECs, reflected by a 'phase-locked'  reproducible interference pattern and near-unity many-realization fringe visibility. Such coherent splitting, maintaining a definite relative phase between the condensates, was demonstrated in atom-chip experiments \cite{Schumm05}.

The effect of interactions on the fringe visibility of coherent preparations, has recently attracted much attention \cite{Greiner02,Schumm05,Jo07,Widera08,PhaseDiffusion,Vardi,Khodorkovsky08,Boukobza09}. Given time to evolve under the influence of inelastic collisions between the atoms, the relative-phase of the separated BECs  disperses because the basis relative-number (Fock) states oscillate with different frequencies \cite{Greiner02,Schumm05,Jo07,Widera08}. This process has come to be known as {\it phase-diffusion} \cite{PhaseDiffusion,Vardi,Khodorkovsky08,Boukobza09}. Its dynamics which is closely related to the Josephson effect in superconductors, can be studied within the two site Bose-Hubbard model  \cite{BJM,Leggett01}, 
\begin{equation}
\label{Ham}
\hat{H}=-J\Lx+U\Lz^2~,
\end{equation}
as a function of the characteristic interaction strength $u=UN/J$, where $J$ and $U$ denote the coupling and collisional interaction energies, respectively \cite{Boukobza09} . The angular momentum operators $\Lx=(\eidag_1 \ei_2+\eidag_2\ei_1)/2$,  $\Ly=(\eidag_1\ei_2-\eidag_2\ei_1)/(2i)$, and $\Lz=(\hn_1 - \hn_2)/2$, are defined in terms of bosonic annihilation and creation operators $\ei_i$, $\eidag_i$ for particles in condensate~${i=1,2}$, with corresponding particle numbers $\hn_{i}=\eidag_{i}\ei_{i}$, satisfying the conservation law $\hn_1+\hn_2=N\equiv 2\ell$. In the extreme strong-interaction Fock regime $u>N^2$, single-particle coherence is lost on a $(U\sqrt{\ell})^{-1}$ timescale regardless of the initial relative phase $\varphi$ \cite{Greiner02}. Coupling the condensates results in phase-locking  \cite{Hofferberth07}. However, the required coupling strength to arrest phase-diffusion depends on $\varphi$ \cite{Boukobza09}. Whereas a $\varphi=0$ phase is locked already in the strong-interaction Josephson regime $1<u<N^2$, a relative phase of $\varphi=\pi$ is only locked in the weak-interaction Rabi regime $u<1$.

In this work we consider the effect of interactions on the fringe-visibility of the initially {\it incoherent} preparation. Instead of the initial coherent state obtained by fast splitting, it is possible to prepare the relative-number squeezed state by slow separation \cite{Jo07,Esteve08,Javanainen99,Grond09}.  We study the buildup of single-particle coherence between such separated condensates, due to the combined effect of interactions and coupling. In the Fock regime, number squeezed states are a good approximation to the system's eigenstates so that no coherence may form. However, throughout the Josephson regime we find that significant coherence may build up, leading to a {\it non-random} phase distribution in a many-realizations interference experiment.  The resulting fringe visibility $g^{(1)}_{12}$ attains a {\it universal} value of $\sim 1/3$ throughout the Josephson regime, in excellent agreement with a semiclassical phase-space picture.  We also study other,  phase-squeezed preparations obtained by unitary rotations of the relative-number state. Such states are encountered for example, in the phase-acquisition stage of Mach-Zendher interferometry with number-squeezed inputs \cite{TFMZ}.  We find that fringe visibility buildup for these preparations takes place in the Rabi-Josephson transition point and that it is sensitive to the initial bi-valued relative-phase.

Below we use for representation the Fock~space basis states $|\ell,m\rangle_\alpha$ where $\alpha=x,y,z$,  which are the joint eigenstates of ${\hat{\bf L}}^2$ and ${\hat L}_\alpha$, with $\ell=N/2$.  We consider the three Fock preparations,
\begin{equation}
|\ell,0\rangle_\alpha=\left({\hat b}_{\alpha1}^\dag{\hat b}_{\alpha2}^\dag\right)^{N/2}\left | {\rm vacuum}\right\rangle
\end{equation}
with $b_{x1,2}=(\ei_1\pm\ei_2)/\sqrt{2}$, $b_{y1,2}=(\ei_1\pm i\ei_2)/\sqrt{2}$, $b_{z1,2}=\ei_{1,2}$. The states $|\ell,m\rangle_{x,y}$ may be obtained by switching the coupling and the bias potential between the wells, inflicting rapid unitary rotations of the Twin-Fock $|\ell,m\rangle_z$ state.

%%%%%%%%%%%%%%%%%%%%%%%
\begin{figure}[t]
\centering
\includegraphics[width=0.50\textwidth]{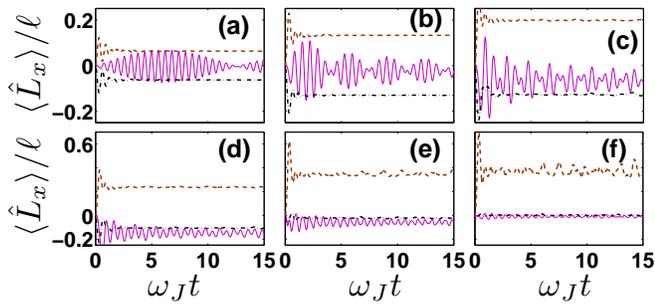}
\caption{(Color online) Evolution of fringe visibility with $N{=}1000$ particles, 
starting from the Fock preparations $|\ell,0\rangle_x$ (solid), $|\ell,0\rangle_y$ (dash-dotted), and $|\ell,0\rangle_z$ (dashed). The coupling parameter is $u=0.5$ (a), $1$ (b) $1.5$ (c), $2$ (d), $10$ (e), and $100$ (f). }
\label{Dynamics}
\end{figure}
%%%%%%%%%%%%%%%%%%%%%%

For all three initial states, the expectation value $\langle\Ly\rangle$, vanishes identically throughout the evolution with the Hamiltonian (\ref{Ham}). Thus, the fringe-visibility in a many-realizations interference experiment is $g^{(1)}_{12}=|\langle\Lx\rangle|/\ell$. In Fig.~\ref{Dynamics} we plot the dynamics of $\langle\Lx\rangle/\ell$ with $N=1000$ particles, starting from the three Fock preparations, for various values of the interaction parameter $u$. We make the following observations: (i) When starting from the site number state $|\ell,0\rangle_z$  interactions lead to the formation of a non-vanishing single-particle coherence when $u<N^2$. This coherence persists well into the Josephson regime. (ii) Similarly, finite coherence is obtained for the $|\ell,0\rangle_{x,y}$ preparations, but only for relatively weak interactions $u\sim 1$. (iii) The coherence evolution for the  $|\ell,0\rangle_{x}$ state exhibits oscillations and beating absent from the dynamics of  the $|\ell,0\rangle_{y}$ preparation, approaching a different mean value.
  
\begin{figure}[t]
\centering
\includegraphics[width=0.5\textwidth]{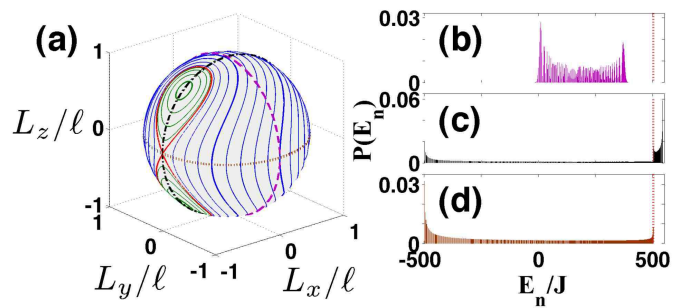}
\caption{(Color online)  Schematic illustration of classical phase-space energy contours for the three Fock preparations $|\ell,0\rangle_x$ (dashed), $|\ell,0\rangle_y$ (dash-dotted), and $|\ell,0\rangle_z$ (dotted), are shown in (a) with $u=1.5$. The corresponding eigenstate occupation probabilities $P_n$  are plotted for $|\ell,0\rangle_x$ (b), $|\ell,0\rangle_y$ (c), and $|\ell,0\rangle_z$ (d). Vertical dotted lines in (b)-(d) correspond to the separatrix energy. Trajectories of maximum tangency (bold) produce caustics in the $P_n$ distribution.}
\label{PSLDOS}
\end{figure}

The dynamics of fringe visibility could be interpreted using a semiclassical picture \cite{Boukobza09,agarwal,Mannschott09,Franzosi00,Graefe07}. Clasical (mean-filed) trajectories are given by the equal energy contours of the Gross-Pitaevskii energy functional,
\begin{eqnarray}  \label{e3}
E(\theta,\varphi) =   
\frac{NJ}{2}\left[\frac{1}{2} u (\cos\theta)^2 - \sin\theta\cos\varphi\right]
\end{eqnarray}
which is restricted to the unit Bloch sphere. In the Rabi regime all motion is essentially linear and the trajectories correspond to slightly perturbed Rabi Oscillations. In the Josephson regime the spherical phase-space is split by a figure-eight separatrix trajectory (Fig~\ref{PSLDOS}(a)), 
to a linear 'sea'  (blue) and two interaction-dominated nonlinear 'islands'  (green).  Finally, in the Fock regime the linear domain becomes smaller than Planck cell, and therefore effectively disappears. In what follows we will assume for simplicity that the interaction is repulsive, i.e.  $u>0$. Since $u\rightarrow -u$, $\varphi\rightarrow \varphi+\pi$ implies $E\rightarrow-E$, the phase space picture for attractive interactions, is a mirror image of the $u>0$ contours.

Semiclassical WKB quantization \cite{Boukobza09,Mannschott09,Franzosi00,Graefe07}  is attained by demanding that ${A(E_n)=(4\pi/N)(n+1/2)}$ where $A(E)$ is the phase-space area enclosed by a fixed energy~$E$ contour, and $4\pi/N$ is the Planck cell. As a result, level spacing is determined from the classical oscillation frequency $\omega(E)=[A'(E)]^{-1}$. The resulting spectrum constitutes in agreement with the classical phase-space structure: (a)  low-energy sea levels extending from ${E_{-}=-\ell J}$ with Josephson-frequency spacing $\omega_J \equiv\omega(E_{-})=\sqrt{(J+NU)J}$, (b) separatrix levels around ${E_{\rm_x}=\ell J}$ with spacing $\omega_{\rm x}=2\omega_J/\log(N^2/u)$, and (c) high-energy island levels, approaching $E_{+} \approx \ell^2U$, with characteristic spacing $\omega_{+} \approx NU$ between nearly degenerate pairs. We note parenthetically that in the strict classical limit ($N\rightarrow\infty$ keeping $u$ fixed) the separatrix frequency vanishes \cite{Wu06,Witthaut06}  but due to the slow $\log(N)$ convergence to classicality \cite{Vardi},  for finite $N$ it only differs from the Josephson frequency by a logarithmic factor. In the Rabi regime, we have $E_+<E_{\rm_x}$ and the entire spectrum consists of sea levels, whereas in the Fock regime the ground state's energy lies above $E_{\rm_x}$ so that all levels reside in the islands.

%%%%%%%%%%%%%%%%%%%%%%%%%
%%%%%%%%%%%%%%%%%%%%%%%%%%

To understand the dynamics of the three Fock preparations, it is useful to consider their Wigner function $\rho(\theta, \varphi)$. The Wigner functions $\rho^{(\alpha)}$ depicting  the $|\ell,0\rangle_\alpha$ preparations, lie along the corresponding great circles around the $\alpha$ axes, as shown in Fig.~\ref{PSLDOS}(a). The Wigner functions of the BHH eigenstates $\rho^{(n)}$ are concentrated along the contour lines of the Gross-Pitaevskii classical energy functional. Thus, the expansion coefficients of the $|\ell,0\rangle_\alpha$ preparation, in terms of the BHH eigenstates $|E_n\rangle$ can be estimated by the semiclassical prescription $P(E_n) = |\langle E_n | \psi\rangle|^2 = \mbox{trace}(\rho^{(n)} \rho^{(\psi)})$. 

In Fig.~\ref{PSLDOS}(b)-(d) we plot the coefficients $P(E_n)$ for the $|\ell,0\rangle_{x,y,z}$ preparations, for $u=1.5$ in the Josephson regime, obtained by direct numerical diagonalization of the BHH. It is clear from  the phase-space landscape of Fig.~\ref{PSLDOS}(a) that the eigenstate expansion should be quite different for the three initial states. The state $|\ell,0\rangle_{x}$ lies entirely in the sea, overlapping a narrow band of linear eigenstates. By contrast, the state $|\ell,0\rangle_{y}$ straddles the entire spectrum, including sea, separatrix, and island levels, whereas the $|\ell,0\rangle_{z}$ state consists of all sea levels up to the separatrix energy, but does not project at all onto the nonlinear islands. This observation is reflected well in the numerical results with the expected extent of the local density of states and caustics obtained for trajectories of  tangency with the initial Wigner distribution. For example, for the $|\ell,0\rangle_{x}$ preparation, the highest expansion coefficients in Fig.~\ref{PSLDOS}(b) are obtained for the energy contours $E_n\approx E(\pi/2,\pi/2)=0$ and $E_n\approx E(\pi,\varphi)=uNJ/4=375 J$ (for $u=1.5$ and $N=1000$), marked by bold (blue) lines in Fig.~\ref{PSLDOS}(a).

%%%%%%%%%%%%%%%%%%%%%%%%%%
\begin{figure}[t]
\centering
\includegraphics[clip,width=0.5\textwidth]{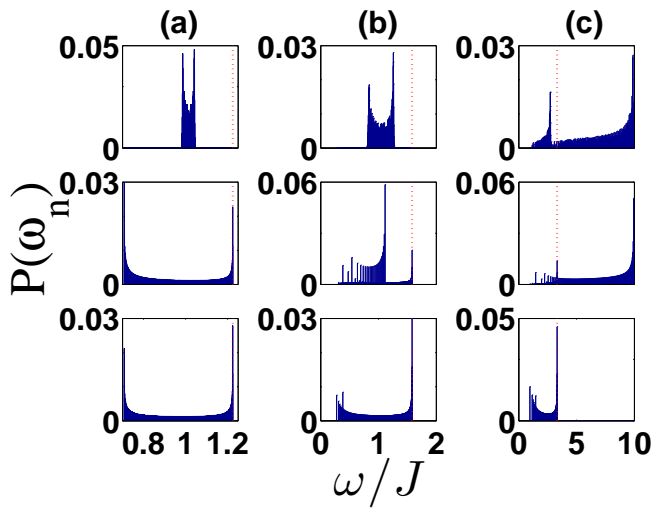}
\caption{(Color online) 
Frequency distributions $\mbox{P}(\omega)$ for the preparations $|\ell,0\rangle_\alpha$, $\alpha=x$ (top), $y$ (middle), and $z$ (bottom) with $N{=}1000$ particles, and $u=0.5$ (a), $1.5$ (b), and $10$ (c).  Dotted lines mark the Josephson frequency $\omega_J$ whereas the low frequency limit corresponds to the separatrix frequency $\omega_{\rm x}$, logarithmically approaching zero with increasing $N$.
}
\label{P_of_omega}
\end{figure}
%%%%%%%%%%%%%%%%%%%%%%%%%%

In order to relate the probability distributions of Fig.~\ref{PSLDOS}(b)-(d) to the time evolution depicted in Fig.~\ref{Dynamics}, we plot the frequency distribution $P(\omega_n)$ with $\omega_n=E_{n+1}-E_n$, for the three initial Fock states, in Fig~\ref{P_of_omega}. In this picture, frequencies extend from the separatrix levels with $\omega\sim \omega_{\mathrm x}$, through the equally-spaced $\omega\sim \omega_J$ low energy sea levels, to the maximal level spacing $\omega\sim \omega_+$ at the top of the nonlinear islands. In the Rabi regime (Fig~\ref{P_of_omega}(a)), the $|\ell,0\rangle_x$ preparation corresponds to a narrow  distribution of sea levels with two dominant frequencies corresponding to the classical trajectories tangential to the $L_x=0$ great circle.  Hence beating is observed around the Josephson frequency. In comparison, the states $|\ell,0\rangle_{y,z}$ have at their disposal the entire sea frequency range, allowing for the continuous buildup of single particle coherence. 

\begin{figure}[t]
\centering
\includegraphics[clip,width=0.50\textwidth]{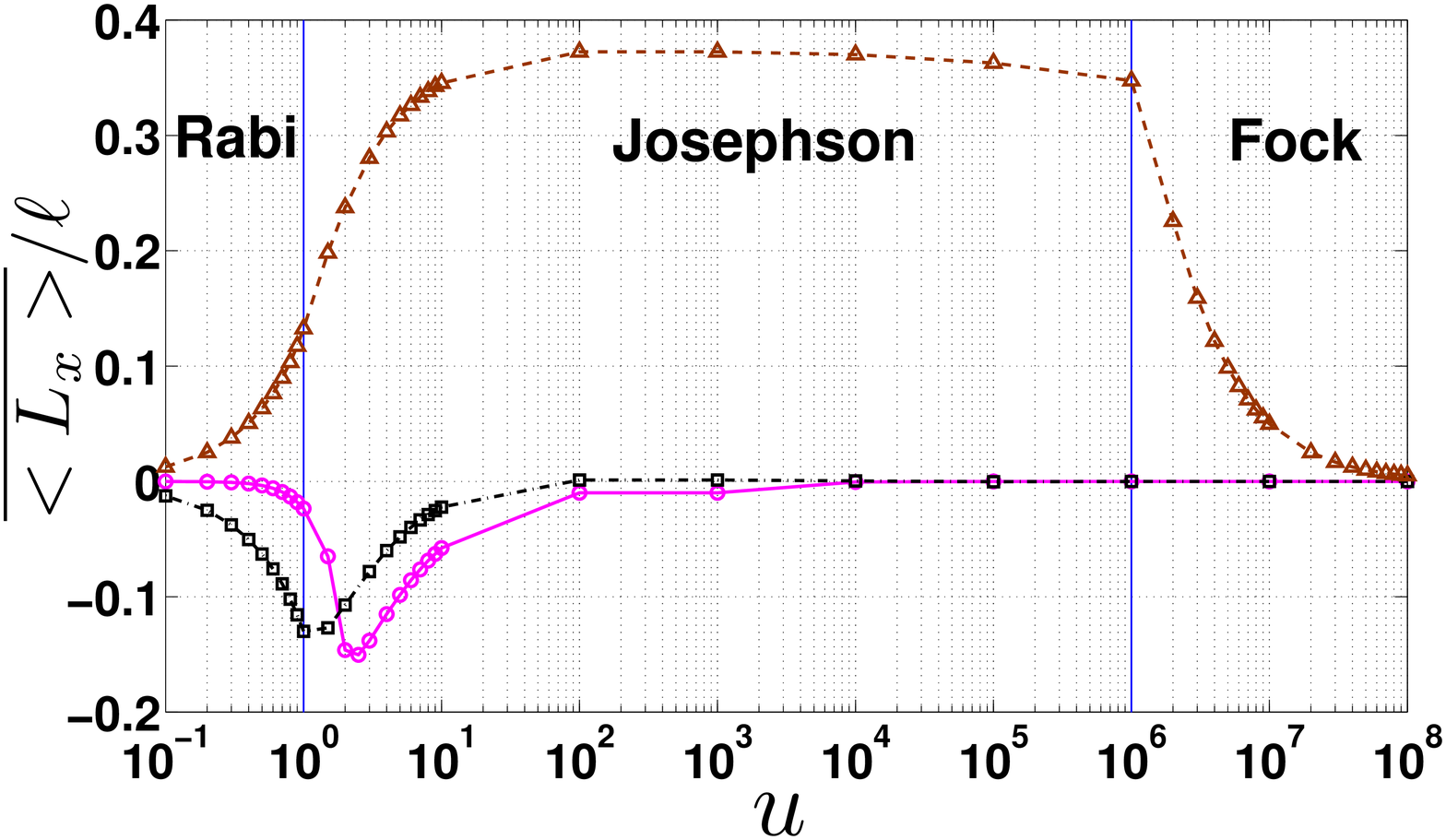}
\caption{(Color online) Acquired time-averaged coherence as a function of the interaction parameter $u$ for the Fock preparations $|\ell,0\rangle_x$ (solid), $|\ell,0\rangle_y$ (dash-dotted), and $|\ell,0\rangle_z$ (dashed).}
\label{Coherence}
\end{figure}

In the Josephson regime (Fig~\ref{P_of_omega}(b),(c)) the separatrix and islands appear. The states $|\ell,0\rangle_x$  and $|\ell,0\rangle_y$ begin to penetrate the islands and project into the high-frequency regime at $u=2$ and $u=1$, respectively. Consequently, if $u$ is large enough, $\langle\Lx\rangle$  never attains a significant magnitude (solid and dash-dotted lines in Fig.~\ref{Coherence}). By contrast, the frequency-span of the preparation $|\ell,0\rangle_z$ remains within the $\omega_{\rm x}<\omega<\omega_J$ range and time-averaged coherence of $g_{12}^{(1)}\sim 0.35-0.40$ at a relative phase $\varphi=0$ is maintained throughout the Josephson regime (Fig.~\ref{Dynamics} and dashed line in Fig.~\ref{Coherence}). Finally, in the Fock regime, the entire phase space is nonlinear,  the $|\ell,0\rangle_z$ preparation becomes a narrow distribution of (island) eigenstates of the BHH, and the interaction does not induce dynamical phase-locking.

%%%%%%%%%%%%%%%%%%%%%%
%%%%%%%%%%%%%%%%%%%%%%

\begin{figure}[t]
\centering
\includegraphics[clip,width=0.50\textwidth]{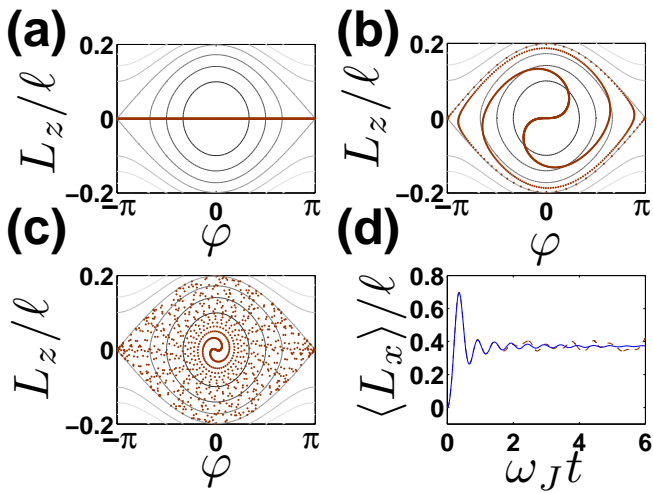}
\caption{(Color online) Popagation of $1001$ classical trajectories with $u=100$, starting from initial conditions corresponding to the Wigner distribution of the twin-Fock preparaion: (a) $t=0$, (b) $\omega_J t=2$, (c) $\omega_J t=50$. Gray lines correspond to the mean-field energy contours of Eq.~(\ref{e3}). The mean value of $L_x$  over all points (solid) is plotted in (d) and compared to the numerical quantum dynamics of Fig.~\ref{Dynamics}(f) (dashed).}
\label{Wigner}
\end{figure}

The value of the interaction-induced fringe-visibility obtained for initially separated condensates can be deduced from the classical time evolution of their Wigner distribution $\rho(\theta,\varphi)$.  The Wigner distribution of the twin Fock state $|\ell,0\rangle_z$ (Fig.~\ref{Wigner}(a)) lies along the $L_z=0$ great circle, passing throughout the linear sea region of phase space, up to the separatrix trajectory. Due to the variation of characteristic frequency from $\omega_{\rm x}$ near $\varphi=\pi$ to $\omega_J$ near $\varphi=0$, propagation in time results in a 'spiral' motion  of the Wigner distribution (Fig.~\ref{Wigner}(b)), which for $\omega_J t\gg 1$ becomes spread out throughout the sea (Fig.~\ref{Wigner}(c)). The phase-distribution, obtained by tracing $\rho(\theta,\varphi)$ over $\theta$ is not uniform since all classical sea trajectories pass through $\varphi=0$ whereas only the separatrix trajectory passes through $\varphi=\pi$. The average value of the coherence $\langle \Lx\rangle$ for this nearly-uniform distribution is given by 
\begin{equation}
\langle \Lx \rangle_{\omega_J t\gg1}
\approx
\frac
{\int_0^\pi \cos\left[\theta_{\rm x}(\varphi)\right]\cos(\varphi)d\varphi}
{\int_0^\pi  \cos\left[\theta_{\rm x}(\varphi)\right](\varphi)d\varphi}
\ell
\label{seaaverage}
\end{equation}
where $\theta_{\rm x}(\varphi)$ is the separatrix energy contour, 
$$\frac{u}{2}\cos^2\left[\theta_{\rm x}(\varphi)\right] -\sin\left[\theta_{\rm x}(\varphi)\right]\cos\varphi=1~.$$
In the Josephson regime we can approximate  $\cos\left[\theta_{\rm x}(\varphi)\right]=\sqrt{2(1+\cos\varphi)/u}$. Substituting this approximate separatrix line into Eq.~(\ref{seaaverage}), we obtain that  $\langle L_x\rangle_{\omega_J t\gg1}=\ell/3$. This value is in good agreement with the numerical results of Fig.~\ref{Coherence}. Averaging numerically over the classical distribution, we find that the coherence dynamics overlaps the numerically-exact quantum calculation (Fig.~\ref{Wigner}(d)), with a quasi-equilibrium value of $\langle L_x\rangle_{\omega_J t\gg1}=0.37\ell$ for $u=100$. This value is {\it universal} throughout the Josephson regime, regardless of the exact values of $u$ and $N$.

%%%%%%%%%%%%%%%%%%%%%%
%%%%%%%%%%%%%%%%%%%%%%

In conclusion, starting from a relative-number state of equally populated BECs (a Twin-Fock state),  interactions in the Josephson regime $1<u<N^2$, result in the build-up of single-particle coherence. This phase-locking process is robust, insensitive to the exact value of the coupling, and has the opposite effect compared to phase-diffusion.  Significant average fringe visibility of $g_{12}^{(1)}\sim 1/3$ is {\it generically} obtained throughout the Josephson regime,  regardless of interaction parameters or particle number , in excellent agreement with semiclassical analysis. Interactions thus serve to select a non-random relative-phase in the weak merging of initially fully-separated condensates. The proposed mechanism is fundamentally different from the deterministic, single-condensate reflection fringes  obtained in collisions of effectively immiscible BECs \cite{Cederbaum07} and requires a much weaker  interaction strength, typical of current BEC interference experiments. For the phase-squeezed states obtained by unitary rotations of the Twin-Fock state, we find phase-sensitive dynamics of the fringe-visibility in the weak interaction regime, also explained to excellent accuracy within the semiclassical picture.

This work was supported  by the Israel Science Foundation (Grant 582/07) and by grant nos. 2006021, 2008141 from the United States-Israel Binational Science Foundation (BSF).
%, and by NSF through a grant for ITAMP at Harvard University and Smithsonian Astrophysical Observatory.

\vspace*{-4mm}

%%%%%%%%%%%%%%%%%%%%%%%%%%%%%%%%%%%%%%%%%%%%%%%%%%%%%%%%%%%%%%%%%%%%%%%%%%%%%%
%%%%%%%%%%%%%%%%%%%%%%%%%%%%%%%%%%%%%%%%%%%%%%%%%%%%%%%%%%%%%%%%%%%%%%%%%%%%%%

%%%%%%%%%%%%%%%%%%%%%%%%%%%%%%%%%%%%%%%%%%%%%%%%%%%%%%%%%%%%%%%%%%%%%%%%%%%%%%
%%%%%%%%%%%%%%%%%%%%%%%%%%%%%%%%%%%%%%%%%%%%%%%%%%%%%%%%%%%%%%%%%%%%%%%%%%%%%%

\end{document}